# Spectroscopic signatures of many-particle energy levels in non-covalently doped single-wall carbon nanotubes

T.V. Eremin[1,2,5], P.A. Obraztsov[2,3], V.A. Velikanov[2,4], T.V. Shubina[5], E.D. Obraztsova[2]

[1] Faculty of Physics of M.V. Lomonosov Moscow State University, Leninskie Gory Str. 1, 119991 Moscow, Russia

[2] A.M. Prokhorov General Physics Institute, RAS, 38 Vavilov street, 119991, Moscow, Russia

[3] Department of Physics and Mathematics, University of Eastern Finland, Joensuu, Finland

[4] National Research Nuclear University, Moscow Engineering Physics Institute, 31 Kashirskoe Highway, 115409 Moscow, Russia

[5] Ioffe institute, RAS, Politechnicheskaya street, 26, 194021 St Petersburg, Russia

**Abstract**

We report the first observation of an optical transition from a ground trion state T to excited trion state T* in (6,5) single-wall carbon nanotubes non-covalently doped with hydrochloric acid. The position of such an excited trion level T* is estimated as 2,12 eV, while the ground trion level T has an energy of 1,08 eV. Besides, pump-probe transient absorption spectroscopy indicates that the ground trion level T cannot be excited directly. Instead, we propose that trions form after nonradioactive relaxation from excitons, dressed by interaction with doping induced hole-polarons. We also report a complete exciton-to-trion conversion by means of photoluminescence spectroscopy, thus supporting existence of the polaron-dressed exciton energy level in p-doped single-wall carbon nanotubes.

**Introduction**

After the first direct experimental confirmation [1] of excitonic nature of optical transitions in single-wall carbon nanotubes (SWNTs), it became clear that multi-particle interactions exert a significant influence on physical properties of SWNTs. Accounting of electon-electon, electron-hole and exciton-phonon interactions successfully explained a set of phenomena, which were inexplicable in frames of simple non-interacting model. The ratio problem, i.e. inequality of the ratio between the second and the first optical transition energies to two, curiously low quantum yield of SWNT photoluminescence and phonon-side bands are among such phenomena [2,3]. Since that studying of many body interactions in carbon nanotubes has become an increasingly actual research field.

Rønnow et al. have theoretically found that an interaction of excitons with electrons and holes in SWNT leads to formation of trions, which are detectable even at room temperature [4]. The first experimental confirmation of those findings was performed by Matsunaga at al. (2011), who observed new peaks in absorption and photoluminescence spectra of SWNT after p-doping [5]. The corresponding energy levels, located approximately 100-200 meV below the bright exciton depending on the nanotube diameter, was attributed to positive trions.

The rise of trion-to-exciton band intensity rate in the photoluminescence spectra with increasing the doping level was reported by several groups [5–9]. Under the high level of doping the intensity of the trion band reached the intensity of the exciton band or even slightly exceed it [10]. That tendency was also accompanied by suppression of total photoluminescence brightness. Similar behavior of trion and exciton spectral features was observed in electroluminescence spectra of electrochemically doped SWNTs [11].

It was also reported that direct pumping of the trion level T in doped SWNTs leads to partial occupancy of the trion energy level, which decreased probability of optical transition to the trion level from the ground one ($Gr \stackrel{rad}{\Longrightarrow} T$ transition) [12]. However, it is claimed in very recent study that a direct excitation of trion ($Gr \stackrel{rad}{\Longrightarrow} T$) has no or very low oscillator strength, and absorption band previously ascribed to trion formation indeed corresponds to excitation of polaron-dressed exciton energy level lying above trion's one ($Gr \stackrel{rad}{\Longrightarrow} {}^pE_{(6,5)}$) [13]. Proving this statement, Bai at al. appeal, inter alia, to observed optical transition $T \stackrel{rad}{\Longrightarrow} T^*$ from ground trion level to excited trion level and a time delay between occupation of ${}^pE_{11}$ and $T$ energy levels in highly homogenous SWNTs doped via polymer-wrapping. A kinetic model developed to fit those experimental findings also predicts possibility of complete conversion ($E_{11} \stackrel{nonrad}{\Longrightarrow} T$) of excitations from exciton to trion energy level.

In this work, we first present an observation of the optical transition $T \stackrel{rad}{\Longrightarrow} T^*$ from the ground trion energy level to the excited trion level in single-walled carbon nanotubes non-covalently doped by acid. We propose, that occupation of trion level occurs not as a result of direct optical excitation from ground level ($Gr \stackrel{rad}{\Longrightarrow} T$), but due to relaxation from the closely lying ${}^pE_{11}$ level (${}^pE_{11} \stackrel{nonrad}{\Longrightarrow} T$), since we observe an indication on a time delay between occupation of ${}^pE_{11}$ and T energy levels. We also demonstrate a complete exciton-to-trion conversion $E_{11} \stackrel{nonrad}{\Longrightarrow} T$ in SWNTs, doped under suitable conditions, thus indirectly supporting an existence of the intermediate energy level ${}^pE_{11}$, lying slightly above trion's one.

**Results and Discussions**

Figure 1a shows photoluminescence spectra of suspensions with different concentration of HCl under resonant excitation of the $E_{22}$ excitonic transition in (6,5) nanotubes (570 nm). The distinct peaks labeled as $E_{11}$, and $E_{11}^{dark}$ are due to a radiative decay of bright exciton of (6,5) nanotubes and a phonon-involving radiative decay of K-momentum dark exciton [14] of (6,5) nanotubes, respectively. An illegible peak around 1025-1030 nm should be assigned to emission from (7,5) nanotubes following an exciton energy transfer [15] from (6,5) nanotube. A peak at 890 nm corresponds to a radiative decay of bright exciton in (6,4) nanotubes.

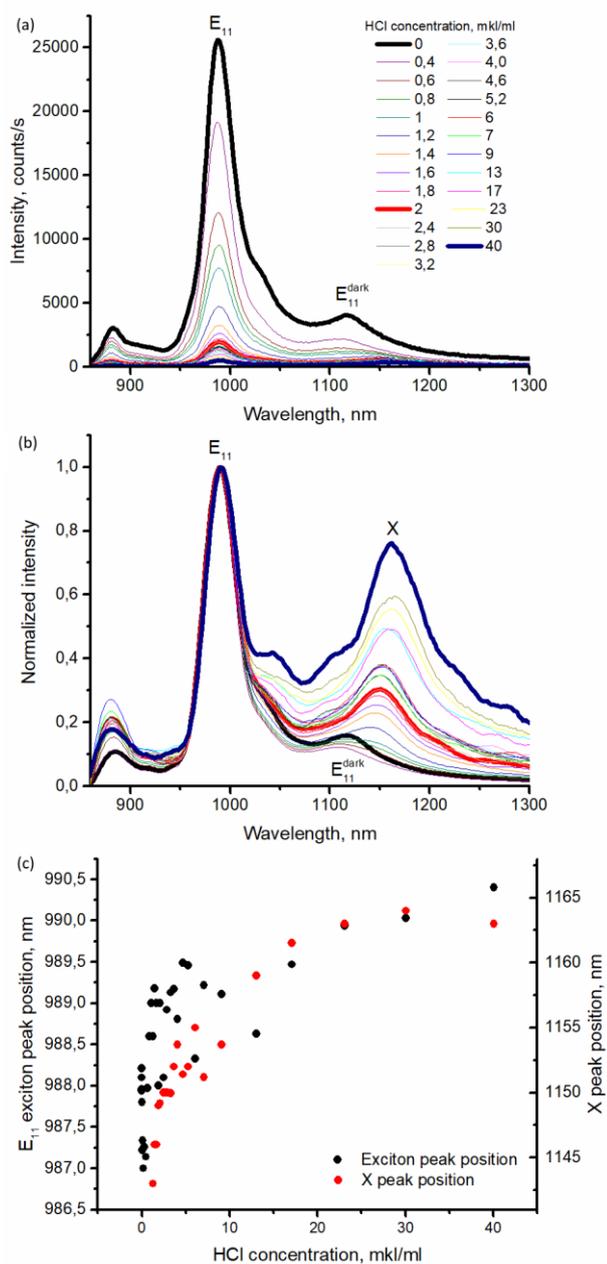

Figure 1 a) The photoluminescence spectra of the suspensions with different concentrations of HCl. The excitation wavelength 570 nm. b) The same as in (a) normalized to the maximum. c) Dependencies of the $E_{11}$ (red circles) and the X (black circles) photoluminescent peak positions on the concentration of HCl in the suspension.

One can also observe another distinct peak labeled as X and rising with increasing concentration of HCl in normalized data plotted in Fig. 1b. Several interpretations of similar red-shifted satellite of the main excitonic peak $E_{11}$ in modified nanotubes are presented in literature. Such peaks observed in nanotubes, covalently doped by aryl diazonium dopants [16], carboxyl groups [17], 4-bromobenzenediazonium tetrafluoroborat and oxygen [18] were ascribed to the excitons localized on defects induced by covalent functionalization of nanotubes. The energy difference between the main exciton and the defect localized exciton noticeably varies depending on the chemical structure of functionalizing group [16]. Another interpretation of photoluminescence peak X in doped SWNTs is a radiative decay of trions which were claimed in photodoped [4,19], gate-doped [7,9], HCl doped [20] and $F_4TCNQ$ doped SWNT [5].

If the X peak were due to the decay of defect-localized excitons, we would observe a rise of defect fingerprints in Raman spectra of nanotubes with increasing concentration of HCl. However, the defect mode (D) intensity does not increase relatively to the tangential mode (G) intensity while the HCl concentration increases (see Fig. 2). Thus, we deny the association of the X peak with defect-localized excitons and, instead of this, explain the X peak as a result of optical decay of trions ($T \stackrel{rad}{\Longrightarrow} Gr$) in doped SWNTs.

Another argument in favor of trion-involving explanation of the X peak is that experimentally measured energy difference (185 meV) between the excitonic peak $E$ and the red-shifted satellite X is close to the theoretically calculated [4] energy difference (199 meV) between the exciton and the trion energy levels.

The 7% discrepancy should be attributed to the environmental effects. Indeed, as shown in Fig.1c, not only intensity, but also the position of X peak is sensitive to hydrochloric acid. Since E and X peaks demonstrate very similar position dependency on HCl concentration, we suggest that the observed red-shift is due to influence of HCl on the local dielectric constant and, consequently, the screening efficiency [21,22].

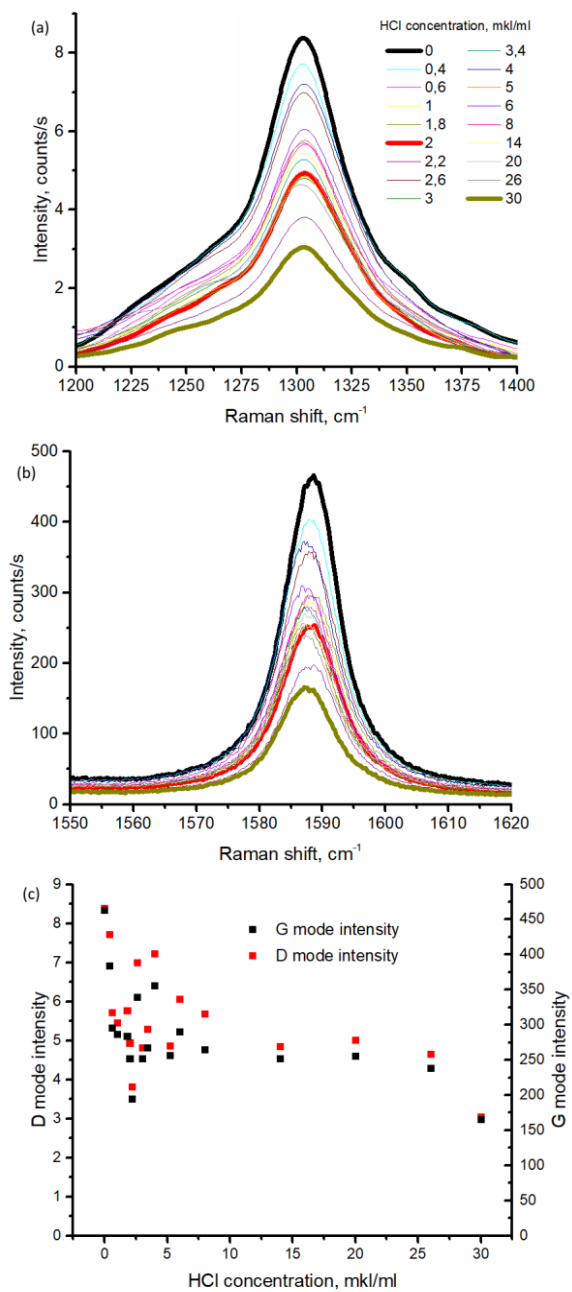

Figure 2 Raman spectra of the suspension with different HCl concentrations in the spectral regions of D mode (a) and G mode (b). c) Dependence of G mode (black squares) and D mode (red scuares) intensities on the concentration of HCl in the suspension.

The suspension with 2 mkl/ml concentration of HCl exhibiting the X photoluminescence peak centered at 1150 nm (solid red line in Fig.1) was investigated by pump-probe technique. Figure 3a shows a dependence of induced optical density of the sample on a time delay between a pump pulse centered at $E_{22}$ resonance of (6,5) nanotubes (570 nm) and a probe pulse covering the spectral region 990-1210 nm. The occupancy of $E_{11}$ excitonic level of (6,5) nanotubes reveals itself as an induced transmittance centered approximately at 990 nm. The vertical cuts of Fig.3a. presented in Fig.3b show changes in the optical density of the sample at several fixed time delays between pump and probe pulses.

We start with discussion of induced absorption in the spectral region around 1190 nm at the longer time delays. This observation is consistent with the results, presented by Bai at al. and strongly supports the idea, that the doping-induced trion level of hole-doped SWNT absorbs light with transition to the excited state ($T \overset{rad}{\Longrightarrow} T^*$). The position of such excited state $T^*$ can be estimated as $2\pi\hbar\left(\frac{1}{1190}+\frac{1}{1150}\right) = 2,12\ eV$, which is approximately 50 meV below the $E_{22}$ excitonic level. That allows us to assume that the excited state T* can correspond to a $E_{22}$ associated trion level.

The ascertaining of physical origin of another observable peak at 1140 nm, labeled as Y, is quite sophisticated. It is often ascribed to a direct optical transition ($Gr \overset{rad}{\Longrightarrow} T$) with excitation of trion in doped SWNT [5,7,12]. However, very recently Bai et al. demonstrated, that such a direct transition does not occur in SWNTs, non-covalently doped via wrapping by [arylene]ethynylene polymer [13]. Bai et al. observed that the maximum population of the trion level (estimated by a transient absorption at $T \stackrel{rad}{\Longrightarrow} T^*$ transition) is reached after a significant time delay after the pump pulse centered at Y peak, thus proving, that the Y absorption band

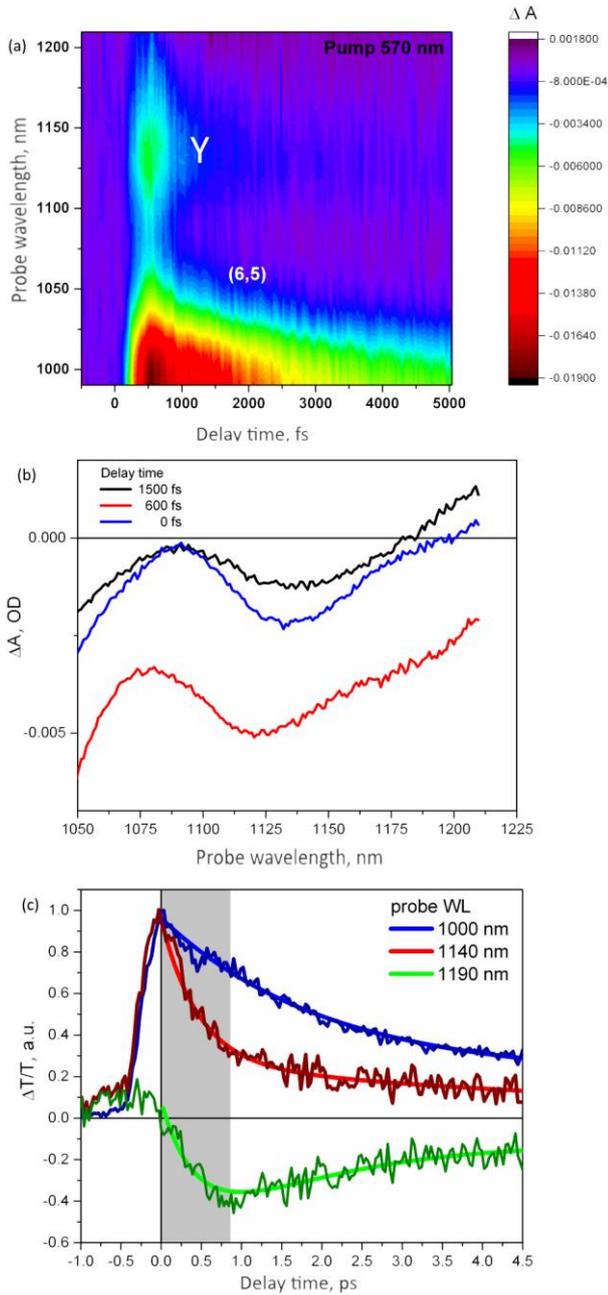

*Figure 3 a) Dependence of induced optical density of the sample on a time delay between a pump pulse centered at $E_{22}$ resonance of (6,5) nanotubes and a probe pulse. b) horizontal cuts of (a). c) vertical cuts of (a).*

corresponds not to the direct excitation of a trion via $Gr \stackrel{rad}{\Longrightarrow} T$ transition, but to the excitation of the polaron-dressed exciton via a $Gr \stackrel{rad}{\Longrightarrow} {}^pE_{11}$ transition. Note, that such a clear evidence was possible due to using of highly chirality- and length- sorted nanotubes. In this work, using much less electrically and morphologically homogenous material, we propose, that Y band does not correspond to $Gr \stackrel{rad}{\Longrightarrow} T$ transition in acid-doped SWNTs as well.

The population dynamics of the energy level, associated with Y spectral feature can be estimated by a horizontal cut of Fig. 3a at 1140 nm, which is presented in Fig. 3c (red line). The population dynamics of the trion energy level $T$ would be estimated by a horizontal cut of Fig. 3a at 1190 nm, but in our experimental conditions the weak transient absorption signal at 1190 nm is significantly overlapped with the strong transient transmittance signal at 1140 nm due to, inter alia, low homogeneity of our material. Thus we estimate the occupancy of the trion energy level by a green line in Fig3c, which is obtained as a horizontal cut of Fig 3a. at 1190 nm ($T \stackrel{rad}{\Longrightarrow} T^*$ band) minus a shoulder contribution of Y band.

It is clearly seen, that the maximum occupation of the trion energy level occurs with a significant time delay after the maximum occupation of the Y band- associated energy level, thus indicating, that the Y band does not correspond to the direct excitation of trions. Following Bai et al. we rather suggest, that the red-shifted satellite Y in absorption spectra of (6,5)-SWNTs, non-covalently p-doped with HCl, corresponds to a $Gr \stackrel{rad}{\Longrightarrow} {}^pE_{11}$ transition with the excitation of exciton ${}^pE_{11}$, dressed by interaction with doping-induced hole-polarons.

Finally, we present a complimentary argument in favor of existence of the ${}^pE_{11}$ level in HCl-doped SWNTs. The kinetic model presented by Bai et al. includes the ${}^pE_{11}$ level and predicts a possibility of a complete exciton-to trion conversion in SWNTs, doped under the certain conditions. If so, at such conditions one would not observe the $E_{11}$ peak in photoluminescence spectrum while observing the bright trion decay peak. We experimentally confirm a possibility of such doping of SWNTs (see photoluminescence spectra in Fig.4). A blue line shows a strong domination of the trion peak in heavily doped SWNTs, while a red line shows the domination of the exciton peak in slightly doped SWNTs. A green line corresponds to the intermediate state with a comparable brightness of exciton and trion peaks in the moderately doped SWNTs. A blue line indicates a complete conversion of photoexcitation from the exciton to the trion energy level, which strongly supports the kinetic model, presented by Bai et al., and indirectly confirms the existence of $^pE_{11}$ level.

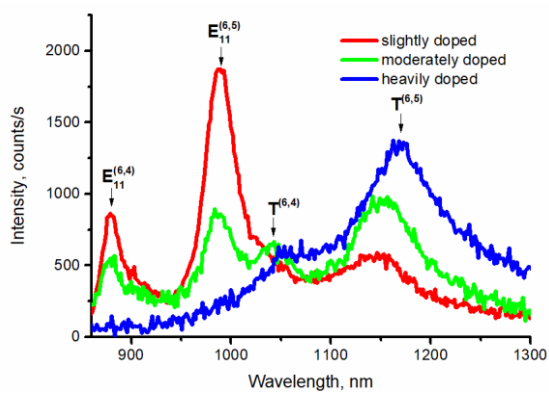

Figure 4 Photoluminescence spectra of slightly (red), moderately (green) and heavily (blue) doped SWNTs.

A schematic illustration of energy levels and nontrivial transitions, discussed in this work is presented in Fig. 5. To summarize, we report a first observation of an optical transition between the ground and excited trion states in SWNTs, non-covalently p-doped by HCl acid. Investigating the relaxation dynamics in morphologically and electrically weakly homogeneous SWNTs, we found, that the absorption band of acid-doped SWNTs, previously ascribed to the direct excitation of trions, is rather corresponds to the excitation of polaron-dressed excitons, which consequently form trions. We also indirectly support such conclusion by observation of compete exciton-to-trion conversion, which was previously predicted in the kinetic model, which includes the polaron-dressed excitons. These findings significantly contribute to understanding of energy structure of doped single-walled carbon nanotubes and many body interactions in low-dimensional materials.

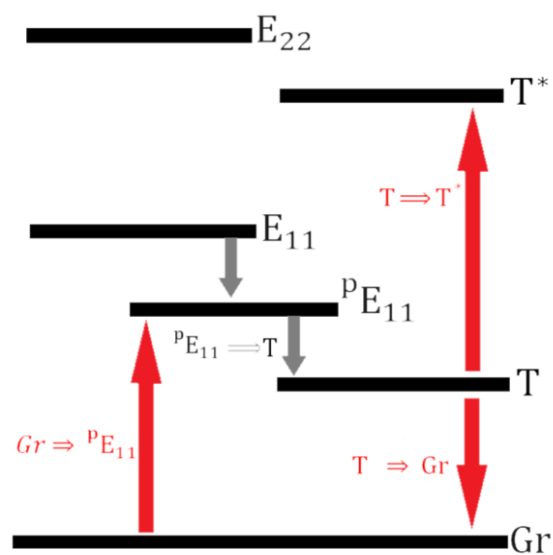

Figure 5 Proposed simplified energy structure of non-covalently doped SWNTs.

### Acknowledgments

This work was supported by RFBR project 17-302-50008. The work was partially supported by Russian Science Foundation grant #17-72-10303 and Academy of Finland Grant #318596

### Materials And Methods

We used (6,5)-enriched CoMoCat single-walled carbon nanotubes powder (Sigma-Aldrich) as a prime material. It was mixed in concentration of 0.1 mg/ml with 2% water solution of SDS. After 4 hours pf tip sonication at room temperature and 1 hour of ultracentrifugation (120000 g), supernatant was collected for further investigations. P-type doping of SWNT was performed by adding of hydrochloric acid in the suspension.

For photoluminescence measurements a "Nanolog-4" spectrophotometer (*Horiba*) was used, where a xenon lamp is an excitation light source, a nitrogen-cooled InGaAs matrix is a detector. Raman spectroscopy was performed using "LabRam" spectrometer (*Horiba*) with a He-Ne laser (633 nm) for excitation and a Peltier-cooled Si detector.

Ti:Sapphire laser radiation passed through OPA amplifier was used for pump probe spectroscopy. For pump pulse, second harmonic was used. Pulse duration was 40 fs for both pump and probe pulses.

## REFERENCES

1. Wang, F., Dukovic, G., Brus, L. E., & Heinz, T. F. (2005). The optical resonances in carbon nanotubes arise from excitons. *Science*, *308*(5723), 838-841.

2. Miyauchi, Y. (2013). Photoluminescence studies on exciton photophysics in carbon nanotubes. *Journal of Materials Chemistry C*, *1*(40), 6499-6521.

3. Dresselhaus, M. S., Dresselhaus, G., Saito, R., & Jorio, A. (2007). Exciton photophysics of carbon nanotubes. *Annu. Rev. Phys. Chem.*, *58*, 719-747.

4. Rønnow, T. F., Pedersen, T. G., & Cornean, H. D. (2010). Correlation and dimensional effects of trions in carbon nanotubes. *Physical Review B*, *81*(20), 205446.

5. Matsunaga R., Matsuda K., Kanemitsu Y. Observation of charged excitons in hole-doped carbon nanotubes using photoluminescence and absorption spectroscopy //Physical review letters. – 2011. – Т. 106. – №. 3. – С. 037404.

6. Park, J. S., Hirana, Y., Mouri, S., Miyauchi, Y., Nakashima, N., & Matsuda, K. (2012). Observation of negative and positive trions in the electrochemically carrier-doped single-walled carbon nanotubes. *Journal of the American Chemical Society*, *134*(35), 14461-14466.

7. Hartleb, H., Späth, F., & Hertel, T. (2015). Evidence for strong electronic correlations in the spectra of gate-doped single-wall carbon nanotubes. *ACS nano*, *9*(10), 10461-10470.

8. Eremin, T., & Obraztsova, E. (2018). Optical Properties of Single-Walled Carbon Nanotubes Doped in Acid Medium. *physica status solidi (b)*, *255*(1).

9. Yoshida, M., Popert, A., & Kato, Y. K. (2016). Gate-voltage induced trions in suspended carbon nanotubes. *Physical Review B*, *93*(4), 041402.

10. Akizuki, N., Iwamura, M., Mouri, S., Miyauchi, Y., Kawasaki, T., Watanabe, H., ... & Matsuda, K. (2014). Nonlinear photoluminescence properties of trions in hole-doped single-walled carbon nanotubes. *Physical Review B*, *89*(19), 195432.

11. Jakubka, F., Grimm, S. B., Zakharko, Y., Gannott, F., & Zaumseil, J. (2014). Trion electroluminescence from semiconducting carbon nanotubes. *ACS nano*, *8*(8), 8477-8486.

12. Nishihara, T., Yamada, Y., Okano, M., & Kanemitsu, Y. (2013). Trion formation and recombination dynamics in hole-doped single-walled carbon nanotubes. *Applied Physics Letters*, *103*(2), 023101.

13. Bai, Y., Olivier, J. H., Bullard, G., Liu, C., & Therien, M. J. (2018). Dynamics of charged excitons in electronically and morphologically homogeneous single-walled carbon nanotubes. *Proceedings of the National Academy of Sciences*, 201712971.

14. Kadria-Vili, Y., Bachilo, S. M., Blackburn, J. L., & Weisman, R. B. (2016). Photoluminescence side band spectroscopy of individual single-walled carbon nanotubes. *The Journal of Physical Chemistry C*, *120*(41), 23898-23904.

15. Ma, Y. Z., Valkunas, L., Dexheimer, S. L., Bachilo, S. M., & Fleming, G. R. (2005). Femtosecond spectroscopy of optical excitations in single-walled carbon nanotubes: Evidence for exciton-exciton annihilation. *Physical review letters*, *94*(15), 157402.

16. He, X., Hartmann, N. F., Ma, X., Kim, Y., Ihly, R., Blackburn, J. L., ... & Tanaka, T. (2017). Tunable room-temperature single-photon emission at telecom wavelengths from sp 3 defects in carbon nanotubes. *Nature Photonics*, *11*(9), 577.

17. Brozena, A. H., Leeds, J. D., Zhang, Y., Fourkas, J. T., & Wang, Y. (2014). Controlled defects in semiconducting carbon nanotubes promote efficient generation and luminescence of trions. *ACS nano*, *8*(5), 4239-4247.

18. Akizuki, N., Aota, S., Mouri, S., Matsuda, K., & Miyauchi, Y. (2015). Efficient near-infrared up-conversion photoluminescence in carbon nanotubes. *Nature communications*, *6*, 8920.

19. Yuma, B., Berciaud, S., Besbas, J., Shaver, J., Santos, S., Ghosh, S., ... & Hönerlage, B. (2013). Biexciton, single carrier, and trion generation dynamics in single-walled carbon nanotubes. *Physical Review B*, *87*(20), 205412.

20. Koyama, T., Shimizu, S., Miyata, Y., Shinohara, H., & Nakamura, A. (2013). Ultrafast formation and decay dynamics of trions in p-doped single-walled carbon nanotubes. *Physical Review B*, *87*(16), 165430.

21. Miyauchi, Y., Saito, R., Sato, K., Ohno, Y., Iwasaki, S., Mizutani, T., ... & Maruyama, S. (2007). Dependence of exciton transition energy of single-walled carbon nanotubes on surrounding dielectric materials. *Chemical physics letters*, *442*(4-6), 394-399.

22. Spataru, C. D., & Léonard, F. (2010). Tunable band gaps and excitons in doped semiconducting carbon nanotubes made possible by acoustic plasmons. *Physical review letters*, *104*(17), 177402.